\documentstyle[preprint,aps]{revtex}

\renewcommand{\ref}[1]{\raisebox{.6ex}{[#1]}}

\newcommand{\be}{\begin{equation}}
\newcommand{\ee}{\end{equation}}

\begin{document}


\title{  Theory of Hall Anomaly in the Mixed State
\footnote{talk given at 1997 Taiwan Superconductivity Conference, Aug. 13-16,
          to appear in Chinese Journal of Physics(Taipei)}
          }

\author{ P. Ao  \\ 
         Department of Theoretical Physics  \\                
         Ume\aa{\ }University, 907 81 Ume\aa, Sweden  }

\maketitle

\begin{abstract}
Analogous to Peierls' arguments for the
`anomalous' Hall in metals I demonstrate that the Hall anomaly 
in the mixed state of superconductors, the sign change of the Hall 
resistivity, is a property of 
a vortex many-body correlation, and 
show that the anomaly is due to the competition between 
vortex vacancies and interstitials.
Within this vortex many-body effect picture, 
many features of the complicated Hall effect
can be understood in quantitative terms. 
For example, 
the expression for the vacancy formation energy is obtained, 
the scaling relation between the Hall 
and longitudinal resistivities with the power varying between 1 and 2 
is found to depend on sample details, 
and near the superconducting transition temperature and 
for small magnetic fields the Hall conductivity is  
proportional to the inverse of the magnetic field 
and to the quadratic of the difference between the measured and 
the transition temperatures.

\end{abstract}



In the past 10 years we have experienced the tremendous successes
of the constructions and applications of vortex many-body 
theories\cite{theory}, such as
in the calculation of the longitudinal resistivity in the mixed state,
a quantity vital for practical applications of superconductors. 
It is somehow surprising to notice that when one encounters 
the transverse resistivity, the Hall effect, 
theoretical models normally proposed 
are mostly of independent vortex dynamics type.
It is now well documented that 
the Hall effect in the mixed state is very rich in behavior. 
Two of its features, the small Hall angle and the
sign change of the Hall resistivity, are very noticeable and are the tests for 
theoretical models. They are usually
referred to as the anomalous Hall effect, which  for short
I will call the Hall anomaly in this paper, although it may
not be a proper name.
Indeed, this  Hall anomaly is ubiquitous 
below the superconducting transition temperature,
occurring in the mixed state of 
both conventional and oxide superconductors.
The crucial ingredient for the Hall effect is the transverse force, 
the Magnus force, on a vortex. 
This force plays the same role as the Lorentz force for electrons. 
Since a straightforward application of the Magnus force in 
any independent vortex dynamics model cannot explain this phenomenon, 
there are questions on the intrinsic nature
of the transverse force in  the vortex dynamic equation.
This force has been subsequently modified into various forms\cite{theory},
which again causes inconsistencies in theories.
Since 60's 
we have the difficult task of constructing a consistent and falsifiable 
Hall anomaly theory.

This situation is identical to what had been
encountered by Peierls in the late 20's
on the usual Hall effect in solids:
It has been known since earlier 20th century that the Hall effect
in metals can be negative, positive, and zero, while
the direct consequence of the Lorentz force in the Drude model
predicts only the negative sign, corresponding to the charge of electrons
in metals. 
This apparent contradiction was the famous anomalous Hall effect at that time.
Peierls resolved this puzzle with the concept of the hole motion,
by consideration of the competition between 
the electron many-body effect caused by Fermi statistics and the
periodic lattice potential. 
Peierls's particle and hole concept has now become a
fundamental building block of modern transport theory in solids, 
widely used in the semiconductor industry.
Similar concept of holes in the filled
electron sea (nowadays called positrons) was conceived by Dirac, 
%
%
and their discovery had won C.D. Anderson the Nobel prize.
%
%
In his classical Hall effect papers\cite{peierls}, 
Peierls stressed that `an explanation 
of this paradox was impossible as long as 
the electrons were visualised as freely-moving in the metal', a statement 
applicable to the present Hall anomaly in the mixed state.

In the present paper using the analogy to the hole Hall effect in solids  
I attempt to solve the Hall anomaly in mixed state by showing 
that it can be understood based on the defect motions, 
vacancies and interstitials, in a pinned vortex lattice, and 
emphasize that the anomaly is a property of the many-body correlation
rather than that of an individual vortex.
There is a complete parallel comparison between
the arguments for the Hall effect in solids and those in the mixed state,
summarized in TABLE I. 
Comparing with the complex mixed state Hall phenomena, 
my theory is undoubtly in its initial state. 
But it appears that the main physics has been captured.  
There are a few quantitatively testable predictions. 
I will demonstrate that vacancies can have the lowest energy scale, 
and they dominate the thermal activation contributions 
to the vortex motion at low temperatures. This causes the sign change of Hall 
effect. The almost equal numbers of vacancies and interstitials, together
with their thermal activation type motions, and guided motions
in the vortex lattice background, make the Hall angle small.
The present theory is consistent with other measurement 
such as the Nernst effect.
In the following I present arguments leading to the theory, 
and discuss its predictions and conditions.
Although the theory 
has only been quantified for a few ideal cases and is necessarily crude 
at this stage,
in the view of Peierls' solution to the older Hall anomaly
I believe it paves the way for a complete theory.  
For simplicity, I will consider an isotropic s-pairing superconductor with
one type of charge carriers in two dimension.
In this situation vortices(or straight vortex lines)
can be viewed as point particles.

 (Put TABLE I here.)

Now I turn to the quantitative side of the present theory.
The vortex dynamic equation for a $j$-th vortex in the sample
takes the form of the Langevin equation identical to that of a
charged particle in the presence of a magnetic field: 
\be
   m \ddot{\bf r}_j = q \frac{\rho_s}{2} h d \; 
     ( {\bf v}_{s,t} - \dot{\bf r}_j )
    \times \hat{z} - \eta \dot{\bf r}_j + {\bf F}_{p} 
    + {\bf f} \; ,
\ee
with an effective mass $m$, 
a pinning force ${\bf F}_{p}$, a vortex viscosity $\eta$, and a
fluctuating force ${\bf f}$. 
In Eq.(1) $q = \pm 1$ is the vorticity, $h$ the Planck
constant, $\rho_s$  the superfluid electron number density at temperature $T$,
$d$ the thickness of the superconductor film, and $\hat{z}$ the unit
vector in $z$-direction.
If there is  a temperature gradient, the thermal force
${\bf F}_T = - s_\phi \nabla T$ should be added in 
at the right hand of Eq.(1),
with $s_\phi$ the entropy carried by a vortex. 
The term associated with the total superconducting electron 
velocity 
${\bf v}_{s,t} = {\bf v}_s + {\bf v}_{s,in}$ 
and the vortex velocity $\dot{\bf r}$ at the right side of Eq.(1)
is the transverse force. 
Although ${\bf v}_{s,t}$ is due to all other vortices, here 
I split it into two parts, with ${\bf v}_s$ corresponding to 
the rearrangement of vortices due to the externally applied supercurrent 
and ${\bf v}_{s,in}$ accounting for the rest contribution describing 
the vortex interaction without external current. 
In the following I will assume ${\bf v}_s$ is small such that 
this splitting is appropriate.
It has been shown that under the general properties of 
a superconductor the transverse force must take such a form as a result of
topological constraint.\cite{ao1}
The large amplitude of this force has been verified by 
a recent experiment\cite{zhu}.

It is a common knowledge that in the mixed state of any real superconductor
the many-body correlation between vortices and the pinning effect
cannot be ignored. The competition between them is the source of the rich 
static and dynamical properties of flux phases\cite{theory}. 
However,
almost all previous models attempted to solve the Hall anomaly have 
ignored this strong many-body correlation, 
which gives rise to the Abrikosov lattice if fluctuations were ignored. 
If there were no pinning for vortices,
the whole vortex lattice would move together under the influence of 
an externally applied current in the same manner as that of 
independent vortices. Hence one would get the same sign of the Hall 
resistivity in both superconducting and normal states.
In the presence of pinning centers
the vortex lattice will be pinned down.
In such a situation the motion of the vortex lattice is made possible 
by various kinds of defect motions caused by thermal fluctuations. 
I will argue below that at low temperatures
vacancies can dominate the contribution to the motion,
and the Hall anomaly occurs.

Let me discuss first the energy scales in the problem.
To simplify the presentation, only thin films will be discussed except
otherwise specified.  
The obvious energy scale will be the depinning of a single vortex.
This energy is $\epsilon_0 \equiv d ( {\Phi_0}/{4\pi \lambda_L } )^2$,
determining by the core energy. 
Here $\Phi_0 = h c/2 e$ is the flux quantum, and 
$\lambda_L^2 = m^{\ast} c^2/8 \pi \rho_s e^2  $  
the London penetration depth, and 
$m^{\ast}$ the effective mass of a Cooper pair.
This energy scale also sets the scale for the 
the creations of a vortex, an antivortex,
and a vortex pair.
One of the simplest collective excitations is a dislocation pair.
Its energy scale is 
$({\epsilon_0}/{ 2\sqrt{3}\pi})\ln({r}/{a_0})$ with $r$ the
distance between dislocations and $a_0$ the vortex lattice constant. 
Apart from a logarithmic factor, 
this energy scale is about 10 times smaller than 
single vortex energy scale $\epsilon_0$.
It is then energetically more favorable to have dislocation pairs
in the lattice comparing.
For temperature $T << \epsilon_0$ I can ignore the 
contribution from the vortices hopping out of pinning centers
and the  creation of vortex-antivortex  pairs.
The vortex lattice is then effectively pinned down.
The smallest energy scale will correspond to point-like defects,
vacancies and interstitials. 
If viewing them as the smallest dislocation pairs, 
I immediately have the estimated energy scale for their formation energies
as, by putting $r \sim 2 a_0$,
\be
    \epsilon_v \sim 
           \frac{1}{ 2\sqrt{3} \pi } 
            \left(\frac{ \Phi_0 }{4\pi \lambda_L} \right)^2 d \; .  
\ee
This result is valid for an intermediate magnetic field $B$: 
$ H_{c1} < B < H_{c2}/2$.  
The linear film thickness $d$ dependence 
of the vacancy formation energy 
$\epsilon_v$ holds for thin enough films,
and independent of the 
thickness for thicker ones due to the $z$-direction correlation, in which 
the straight line assumption in reaching Eq.(2) is no longer valid.
I do not have a quantitative theory for the crossover thickness yet, but
in thin enough films Eq.(2) has been verified.\cite{ao2}
It is clear from the above analysis
that vacancies and interstitials have the lowest excitation energy scales.

Based on the general observations for the defect formations in crystals and 
the theoretical and experimental evidences in vortex lattices 
I argue now that the vacancy formation energy 
is even lower than that of an interstitial for a wide parameter regime.
The experimental observations at low magnetic fields have shown
the abundance of vacancies comparing with interstitials.
The natural explanation is that the
the vacancy formation energy is lower than that of interstitials,
therefore by thermal fluctuations vacancies have a higher density.
The theoretical understanding is that the lower vacancy formation energy is  
a result of the strong short range and repulsive nature of the 
interaction between vortices at low magnetic fields, as have been 
shown.
Similar phenomenon has also been observed in other crystalline 
structures. 
At high magnetic fields the vortex interaction becomes of long range 
comparing with the vortex lattice constant.
The short range behavior which determines the formation energy difference  
between vacancies  and interstitials becomes relatively 
unimportant.
The difference between formation energies of vacancy and interstitial will 
be smaller in this limit.
Overall, vacancies  have a lower formation 
energy for a finite range interaction.  
I conclude that they will dominate thermal fluctuation contributions 
to resistivities at low enough temperatures.\cite{ao2}

Using the analogy to the dynamics of a hole or a particle
in a solid in the presence of a magnetic field,  with 
a pinned perfect vortex lattice as a filled valence band and a vacancy in real
space as a hole in the energy space, illustrated in the TABLE I, 
the desired transverse force on a defect, vacancy or interstitial,
can be obtained as
\be
   {\bf F}_M^d = \mp q \frac{\rho_s}{2} h d \; 
     ({\bf v}_s - \dot{\bf r}_0 ) \times \hat{z} \; .
\ee
This equation shows that both a vacancy and an interstitial will
move along the direction of the applied supercurrent ${\bf v}_s$.
This implies that vortices defining vacancies move against the direction of 
${\bf v}_s$, a result of the many-body correlation and pinning.
This leads me to the main conclusion that at low enough 
temperatures the sign of the Hall resistivity is different from its sign in 
the normal state because of the dominance of vacancies.
Quantitatively, vacancies and interstitials 
may be considered as independent particles moving in the
periodic potential formed by the vortex lattice and a random potential
due to the residue effect of pinnings.
The potential height of the periodic potential as well as that of 
the random potential is an order
of $\epsilon_v$. 
Assuming the vacancy (interstitial) density $n_{v}(n_{i})$ 
in a steady state, the longitudinal resistivity is
\be
    \rho_{xx} = \frac{h}{2 e^2} \sum_{l=v,i}
             \frac{\eta_{l} \; {\rho_s d h }/{2} }
                      { \eta_{l}^2 + ({\rho_s d h }/{2})^2 } \;
                \frac{ n_{l} }{\rho_s }  \; ,
\ee
and the Hall resistivity 
\be
   \rho_{yx} =  \frac{ h}{2 e^2} 
            \sum_{l=v,i} q_l \frac{ ({\rho_{s}dh}/{2})^2 }
                   { \eta_{l}^2 + ({\rho_s dh}/{2})^2 } \;
                \frac{ n_{l} }{\rho_s}   \; , 
\ee
with $q_v = - q$ and $ q_i = q$.
Here $\eta_{v,i}$ are the effective vacancy and interstitial viscosities,
related to their diffusion constants in the
periodic potential due to the vortex lattice by the Einstein relation between
the diffusion constant and the mobility.
It should be pointed out that  contributions of other
vortex motions to resistivities such as vortex-antivortex pairs, 
which are omitted here for their 
smaller activation probabilities, 
are additive to those of vacancies, 
and that the including of the normal
fluid (quasiparticle) contributions is straightforward\cite{ao4}.

Under the driving of a temperature gradient 
the effective thermal force felt by a vacancy
is opposite in sign to the force felt by an interstitial 
or a vortex in direction 
but equal in magnitude, ${\bf F}_T^v = + s_\phi \nabla T$.
Then the Nernst effect due to vacancies has the same sign as that of 
vortices or interstitials. Therefore my theory gives that
in the Hall anomaly regime there is no sign change for 
the Nernst effect, and furthermore, 
the Nernst effect is more pronounced because of the
large contribution due to both vacancies and interstitials.
This is in agreement with the experimental 
observations\cite{ao4}.

There are several qualitative implications of the present theory worth
further discussions.
In the above physical picture, to obtain a maximum contribution
of vacancies, I need  the vortex lattice to define vacancies and
a sufficiently strong 
pinnings to prevent the sliding of vortex lattice 
to obtain 
a maximum contribution of vacancies. 
The existence of a whole lattice structure is nevertheless unnecessary.
Sufficiently large local crystalline structures, 
like lattice domains, will be enough to 
define vacancies. Therefore vacancy-like excitations in
a vortex liquid state can exist, 
because of the existence of large local orderings. 
Whether or not this is also true for a vortex glass state 
depending on details. 
For example, a further lowering of temperature may quench a 
vortex system into a glass state with no local crystalline 
structure. The usual counting of defects does not work.
However, a general classification based on the Delauney triangulation method
can still be applied\cite{yu}, and the careful analysis of the defect 
stabilities and
relative energy scales as well as their mobilities remains to be done.
It is possible that the sign of the Hall resistivity will change again
in this parameter regime. 
On the other hand, for a fixed temperature 
if the pinning is too strong, for example, the (random) pinning center density
is much larger than the vortex density, vortices will
be individually pinned down and the local 
lattice structure required 
for the formations of  vacancies and interstitials will be lost. 
This suggests that the Hall anomaly only exists
in a suitable range of pinnings and magnetic fields, that is,
for $B_l < |B| < B_u$ with the lower and upper critical fields 
determined by pinning as well as by temperature. 

To make a contact with experiments 
I work out several limiting cases of Eqs.(4,5) below.
At low temperatures the 
motions of vacancies and interstitials in the vortex lattice 
are thermal hoppings:
 $\eta_{v,i} = \eta_0 \; e^{a_{v,i} \; \epsilon_v /K_B T } $ and
$n_{v,i} = n_0 \; e^{b_{v,i} \; \epsilon_v /K_B T } $.
If $\eta_v,\eta_i \ll \rho_s d h /2$, corresponding to the Hall angle 
$|\tan\theta| = |\rho_{yx}/\rho_{xx} | \ll 1$ 
common in experiments,
\[
   \tan\theta = - q \frac{\rho_s d h }{ 2\eta_0}
      \frac{ e^{ -(2 a_v + b_v) \frac{\epsilon_v}{k_B T} }
           - e^{ -(2 a_i + b_i) \frac{\epsilon_v}{k_B T} } }
           { e^{-(a_v + b_v) \frac{\epsilon_v}{k_B T} }
           + e^{-(a_i + b_i) \frac{\epsilon_v}{k_B T} }    } \: .
\]
For two temperature limits,  
\be
  \tan\theta  = \left\{ \begin{array}{ll}
      - q \frac{\rho_s d h}{2\eta_0}
    \frac{\gamma}{2} \frac{\epsilon_v}{k_B T} \;, & 
     k_B T \geq \epsilon_v \; . \\  
      - q \frac{\rho_s d h}{2\eta_0}
      e^{- a_v \frac{\epsilon_v}{k_B T} } ,
      & k_B T < min\{1,\gamma\}\epsilon_v .\end{array} \right.
\ee  
Here $\gamma = 2 a_i + b_i - 2 a_v - b_v$ is a numerical factor.
The high temperature limit $k_B T \geq \epsilon_v$ is achieved near 
superconducting transition temperature $T_{c0}$, but the thermal creation 
of a vortex-antivortex pair is still relatively
improbable, because the relevant energy
scale $\epsilon_0$ is about 10 times bigger than $\epsilon_v$.
In the low temperature limit 
both longitudinal and Hall resistivities vanish exponentially. 
A scaling relation between them is
\be
   \rho_{yx} = A \; \rho_{xx}^{\nu } \, \,  , {\ } 1 \leq \nu \leq 2
\ee
with $A = - q (\rho_s dh/2\eta_0)^{b_v/(a_v + b_v)} 
          (2e^2 \rho_s/h n_0 )^{a_v/(a_v + b_v)}  $,  
and the power  $\nu = (2 a_v + b_v)/(a_v + b_v)$
varying between 1 and 2, depending on the detail of a sample which determines 
the numerical factors $a_v$ and $b_v$. 
For example, if all vacancies are produced by pinnings, 
$b_v = 0$ and $\nu = 2$.
In the other limit, if all vacancies are produced by thermal activations, 
and if $ a_v << 1$, $b_v = 1$ and $\nu \simeq 1$. 

Another useful quantity is the Hall conductivity 
$\sigma_{xy} = \rho_{yx} /( \rho_{xx}^2 + \rho_{yx}^2 )$.
Under the same assumption of $\eta_v,\eta_i \ll \rho_s d h /2$
I obtain the Hall conductivity due to vacancies and interstitials, 
from Eqs.(4,5), as
$$
  \sigma_{xy} =  - q \frac{ 2e^2 }{h }\frac{ \rho_s }{n_0 }
                \frac{ e^{ -(2 a_v + b_v) \frac{\epsilon_v}{k_B T} }
                     - e^{ -(2 a_i + b_i) \frac{\epsilon_v}{k_B T} } }
                     { \left[ e^{-(a_v + b_v) \frac{\epsilon_v}{k_B T} }
                + e^{-(a_i + b_i) \frac{\epsilon_v}{k_B T} } \right]^2 } \; .
$$
Again, in two temperature limits,
\be
      \sigma_{xy}   = \left\{ \begin{array}{ll}
      - q \frac{ 2e^2 }{h }\frac{ \rho_s }{n_0 }
    \frac{ \gamma }{4} \frac{\epsilon_v}{k_B T} \;, & 
     k_B T \geq \epsilon_v \; . \\  
      - q \frac{ 2e^2 }{h }\frac{ \rho_s }{n_0 }
                e^{ + b_v \frac{\epsilon_v}{k_B T} } ,
      & k_B T < min\{1,\gamma\}\epsilon_v . \end{array}\right. 
\ee    
Here $0 \leq b_v \leq 1$ and 
$\gamma \sim O(1)$.
Near the superconducting transition temperature $T_{c0}$, $\rho_s 
= \rho_{s0} ( 1 - T/T_{c0})$ and $\epsilon_v = \epsilon_{v0}( 1 - T/T_{c0})$
because of the London penetration depth in Eq.(2).
Making a further assumption of $n_0 = B/\Phi_0$, 
with $\Phi_0$ the flux quantum,
from Eq.(8) I obtain  
\be  
   \sigma_{xy} = \alpha_1 \frac{ ( 1 - {T}/{T_{c0} })^2 }{B} \; ,   
\ee
with 
$ \alpha_1 = - q ({ 2e^2 }/{h }) \rho_{s0} \Phi_0 
    \gamma {\epsilon_{v0} }/{4 k_B T_{c0}} $.
Taking $\rho_{s0} = 10^{21} /cm^3$,
$\gamma = 1$, and ${\epsilon_{v0} }/{k_B T_{c0} } = 50$,
$|\alpha_1 |\sim 20 \, T\mu\Omega^{-1}cm^{-1}$.

Two comments are in order.
1. Since vortex interaction terms cancel each other when summing 
over all vortices, one may conclude  that there is no
many-body correlation effect on the sign of the Hall resistivity.
This absence of Hall anomaly is the result of  the
underestimation of the many-body correlation, when applied to electrons
directly contradict to both Peierls' theory and experiments.
2. One may resort the motion of an antivortex in a vortex-antivortex pair 
to explain the anomaly. While it is a physically possible scenario, 
as discussed above
the creation energy of an antivortex is about 10 times larger 
than that of a vacancy, 
which makes it energetically unfavorable at a relatively low temperature. 
Furthermore, since
an antivortex feels the same thermal force as a vortex,
${\bf F}_T = - s_\phi \nabla T$, it has an opposite sign  contribution to 
that of a vortex for the Nernst effect, in conflicting with 
experiments.\cite{ao2,ao4}

Finally, let me summarize the main features of the present theory.
Like the same effect in solids,  Hall effect in the mixed state is 
a many-body phenomenon.
It is impossible to give it a consistent explanation  
by any independent vortex dynamics model, as long ago pointed out
by Peierls in a closely related situation.
The time reverse symmetry of vortex dynamics
is broken by the transverse force, or the Magnus force.
Together with the broken vortex vacancy-interstitial symmetry finite
Hall effect appears.
The dominant role played by defects shows  that
fluctuations are essential in the present theory. 
Although it has several quantitative predictions to be tested, 
the decisive one
may be the combination of the real space-time observation of the 
vortex vacancies and the Hall measurement.

\noindent
{\bf Acknowledgments} 

 This work was supported by Swedish Natural Science Research Council.


\newpage

\Large

\begin{center}
  {  TABLE I. Qualitative Physical Picture }
\end{center}

\normalsize

\begin{center}
    two major conclusions: \\
    1. Many-body correlation responsible for the Hall effect; \\
    2. No one-one correspondence between the Hall effect 
       and the transverse force.
\end{center}

\begin{center}
\begin{tabular}{|c|c|c|}          \hline
   &  MIXED STATE     & SOLIDS \\       \hline       
Fundamental particles  
      & Vortices                     &  Electrons            \\ \hline
Many-body ground states 
      & Pinned vortex lattice        &  Filled valence band  \\ \hline
Excitations, I   & Vortex interstitials         &  Particles    \\ \hline
Excitations, II  & Vortex vacancies             &  Holes        \\ \hline
Symmetry condition, I 
     & Interstitial-vacancy {\bf a}symmetry 
              & Particle-hole {\bf a}symmetry \\ \hline
Symmetry condition, II
    & Transverse force(Magnus force) & Lorentz force  \\ \hline
\end{tabular}
\end{center}

\end{document}